\documentclass[
aps,
reprint,
prl,
superscriptaddress,
amsmath,
amssymb,
floatfix,
longbibliography,
noeprint
]{revtex4-2}

\usepackage{bm}
\usepackage{graphicx}
\usepackage{latexsym}
\usepackage{array}
\usepackage{amsfonts}
\usepackage[free-standing-units]{siunitx}
\usepackage[dvipsnames]{xcolor}
\usepackage{comment}

\usepackage{mathtools}  
\usepackage{makecell}   
\usepackage{hhline}     
\usepackage{amsthm}
\usepackage{amscd}

\usepackage{xcolor}
\renewcommand\vec{\boldsymbol}

\definecolor{orange}{rgb}{1,0.5,0}
\definecolor{goodGreen's}{rgb}{0.1,0.5,0}
\definecolor{goodred}{rgb}{0.7,0,0}

\usepackage[colorlinks,urlcolor=goodGreen's,citecolor=blue,linkcolor=goodred]{hyperref}

\newcommand{\orcid}[1]{\href{https://orcid.org/#1}{\includegraphics[width=8pt]{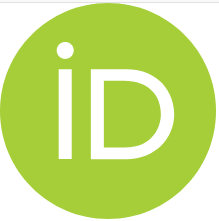}}}

\usepackage{verbatim}
\usepackage[normalem]{ulem}
\graphicspath{{images/}}

\makeatletter
\def\@url#1{}
\makeatother

\begin{document}

\title{Superconductivity as a Probe of Altermagnetism: Critical Temperature, Field, and Current}

\author{A. A. Mazanik \orcid{0000-0001-6389-8653} }
\email{andrei.mazanik@csic.es}
\affiliation{Centro de Física de Materiales (CFM-MPC) Centro Mixto CSIC-UPV/EHU,
E-20018 Donostia-San Sebastián, Spain}

\author{Rodrigo de las Heras \orcid{0009-0003-3640-1498}}
\affiliation{Centro de Física de Materiales (CFM-MPC) Centro Mixto CSIC-UPV/EHU, E-20018 Donostia-San Sebastián,  Spain}

\author{F. S. Bergeret \orcid{0000-0001-6007-4878}}
\affiliation{Centro de Física de Materiales (CFM-MPC) Centro Mixto CSIC-UPV/EHU, E-20018 Donostia-San Sebastián, Spain}
\affiliation{Donostia International Physics Center (DIPC), 20018 Donostia-San Sebastian, Spain}

\date{\today}

\begin{abstract}
We study thin films that host coexisting collinear $d$-wave altermagnetic and superconducting orders in the presence of an external magnetic field that fully penetrates the films. We use the Ginzburg–Landau functional  to analyze the response of the films to magnetic fields and an in-plane supercurrent. We demonstrate that the interplay between superconductivity and altermagnetism induces characteristic fourfold anisotropies in the critical temperature, parallel critical field, and critical current density of the films. These results establish experimentally accessible signatures of altermagnetism in superconducting films and in superconductor/altermagnetic insulator heterostructures.
\end{abstract}

\maketitle

A new form of magnetic order, altermagnetism, has recently been predicted \cite{Smejkal2022Emerging,Smejkal2022Beyond,Tamang2025Review}. Altermagnets exhibit spin-split electronic bands in momentum space while maintaining zero net magnetization. The combination of these two properties, distinct from both ferromagnets and conventional antiferromagnets, establishes altermagnetism as a fundamentally new magnetic phase with promising implications for spin-dependent transport and ultrafast spintronics. Consequently, developing reliable experimental and theoretical diagnostics capable of unambiguously distinguishing the altermagnetic order from conventional antiferromagnetism is essential for both its fundamental classification and the realization of altermagnet-based functionalities.
Possible experimental probes include angle-resolved photoemission spectroscopy \cite{Lee2024_ARPES_MnTe,Reimers2024,Yang2025}, transport measurements based on the anomalous Hall effect and nonlinear transport responses \cite{Sinova2024_Mn5Si3,Song_2025_Hall,Mali2026}, and thermal transport measurements \cite{Sinova2024_Nernst,Badura2025}. All of these techniques are experimentally demanding and require either sophisticated measurement setups or careful analysis of charge and heat transport data, typically involving the disentanglement of contributions from different microscopic mechanisms, symmetry considerations, and first-principles calculations tailored to a specific material.

\begin{figure}[htbp]
    \centering
    \includegraphics[width=0.8\linewidth]{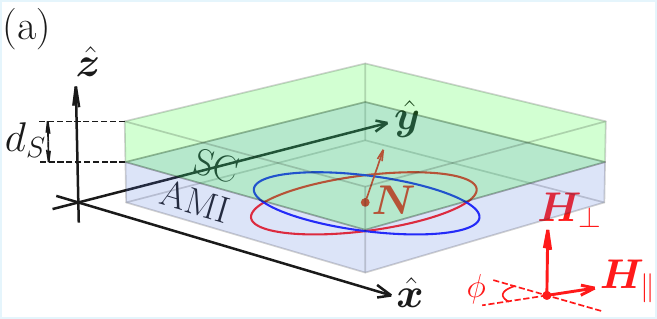}
    \includegraphics[width=0.49\linewidth]{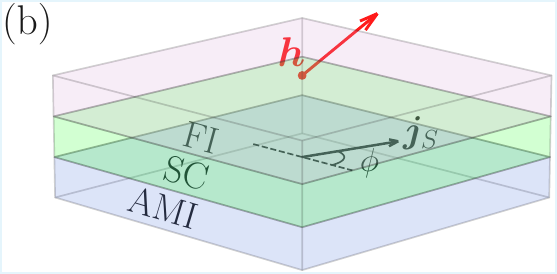}
    \includegraphics[width=0.49\linewidth]{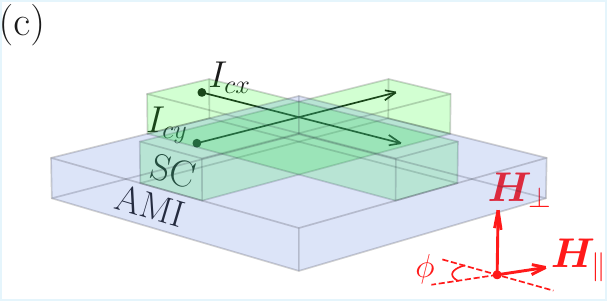}
    \caption{
    (a): Altermagnetic superconducting film realized as a bilayer composed of a conventional $s$-wave superconductor (SC) and an altermagnetic insulator (AMI). 
Although the Néel vector $\vec{N}$ may point in an arbitrary direction, the spin splitting occurs in the $\hat{\vec{x}}$--$\hat{\vec{y}}$ plane, resulting in an altermagnetic tensor $K_{jk}$ with the only two non-vanishing components which we set as $K_{xx} = -K_{yy} = K$. 
The film is subjected to the external magnetic field $\vec{H} = (H_\parallel \cos\phi, H_\parallel \sin\phi, H_\perp)$. 
(b): Ferromagnetic insulator/SC/AMI trilayer. The ferromagnetic insulator induces an exchange field $\vec{h}$ in the superconductor coupled to the altermagnetic Néel vector $\vec{N}$ given by the AMI film. 
(c): Cross-shaped SC/AMI structure suitable for detecting the critical-current anisotropy $I_{cx} \neq I_{cy}$ induced by the external magnetic field $\vec{H}$.  }
    \label{fig:setup}
\end{figure}

Shortly after altermagnetic order was predicted, the idea of combining altermagnets with superconductors has been explored. In particular, the coexistence of superconductivity and altermagnetism, either within a single material or in heterostructures composed of altermagnets (or altermagnetic insulators) and superconductors, has been studied in different works \cite{Zyuzin2024,Linder2024_Memory,Belzig_2025_Thermodynamic,
Annica_BS2025_Perfect_Diode,
AnnicaBS2025_Constraints,
Kokkeler_2025,
Knolle_2025_PDW_SC_diode,
bobkov2026inverse,
sachin2026altermagnetic,
schrade2026altermagnetic1,
Linder_2023_Josephson_Altermagnets,
Tanaka2024_AM_JJ,
heras2025interplay,
Justin2025_magnetoelectric,kokkeler2025nonequilibrium,vasiakin2025disorder}. In such systems, altermagnetism modifies the electronic and spin-dependent transport properties, manifesting itself in Josephson and superconducting diode effects \cite{Linder_2023_Josephson_Altermagnets,Tanaka2024_AM_JJ,Annica_BS2025_Perfect_Diode,Knolle_2025_PDW_SC_diode,heras2025interplay,sachin2026altermagnetic,schrade2026altermagnetic1}, magnetoelectric responses \cite{Zyuzin2024,Kokkeler_2025,
Justin2025_magnetoelectric,heras2025interplay}, and nonequilibrium spin-splitter effects \cite{kokkeler2025nonequilibrium}.

\begin{figure}
    \centering
    \includegraphics[width=0.92\linewidth]{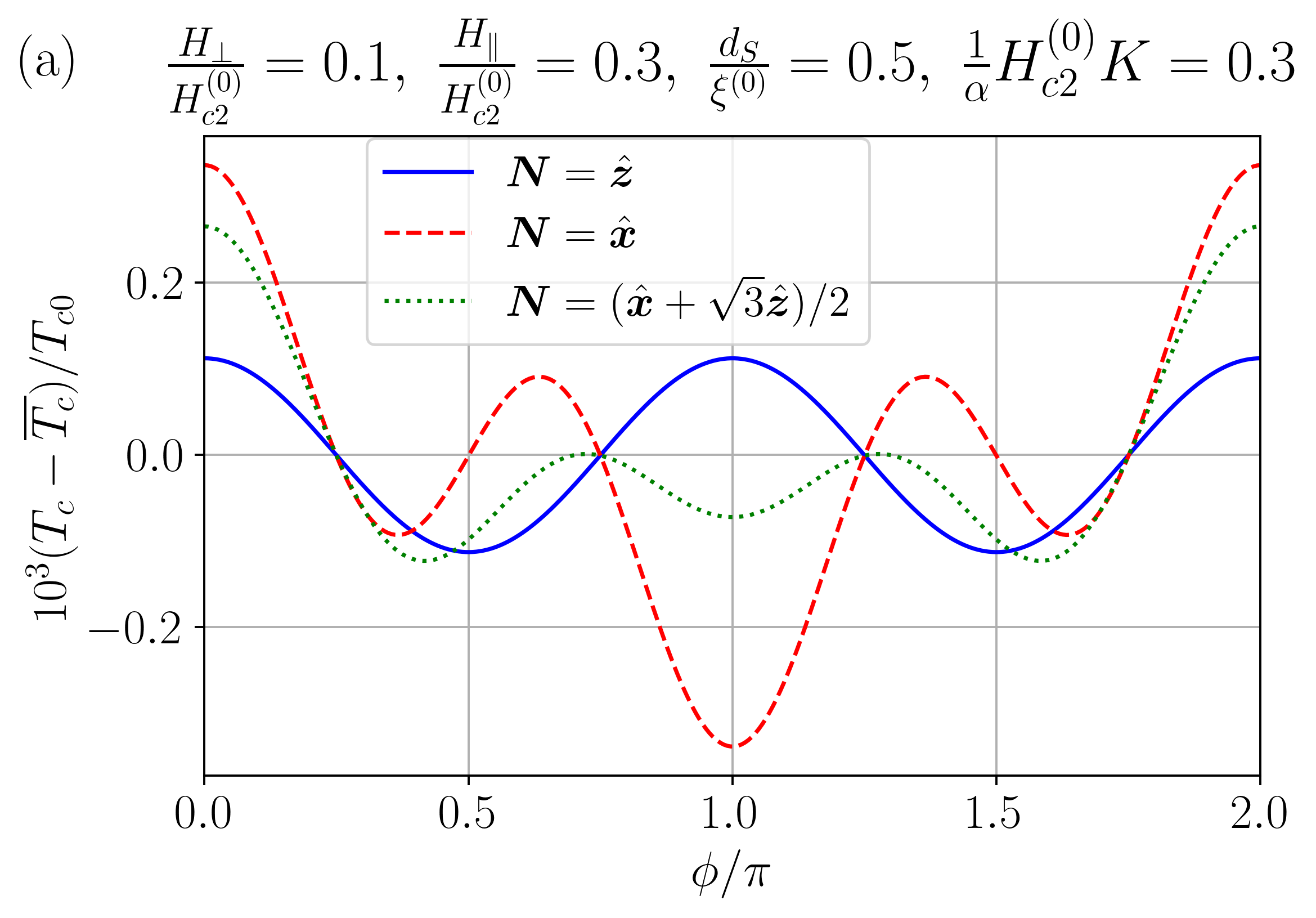}
    \includegraphics[width=0.92\linewidth]{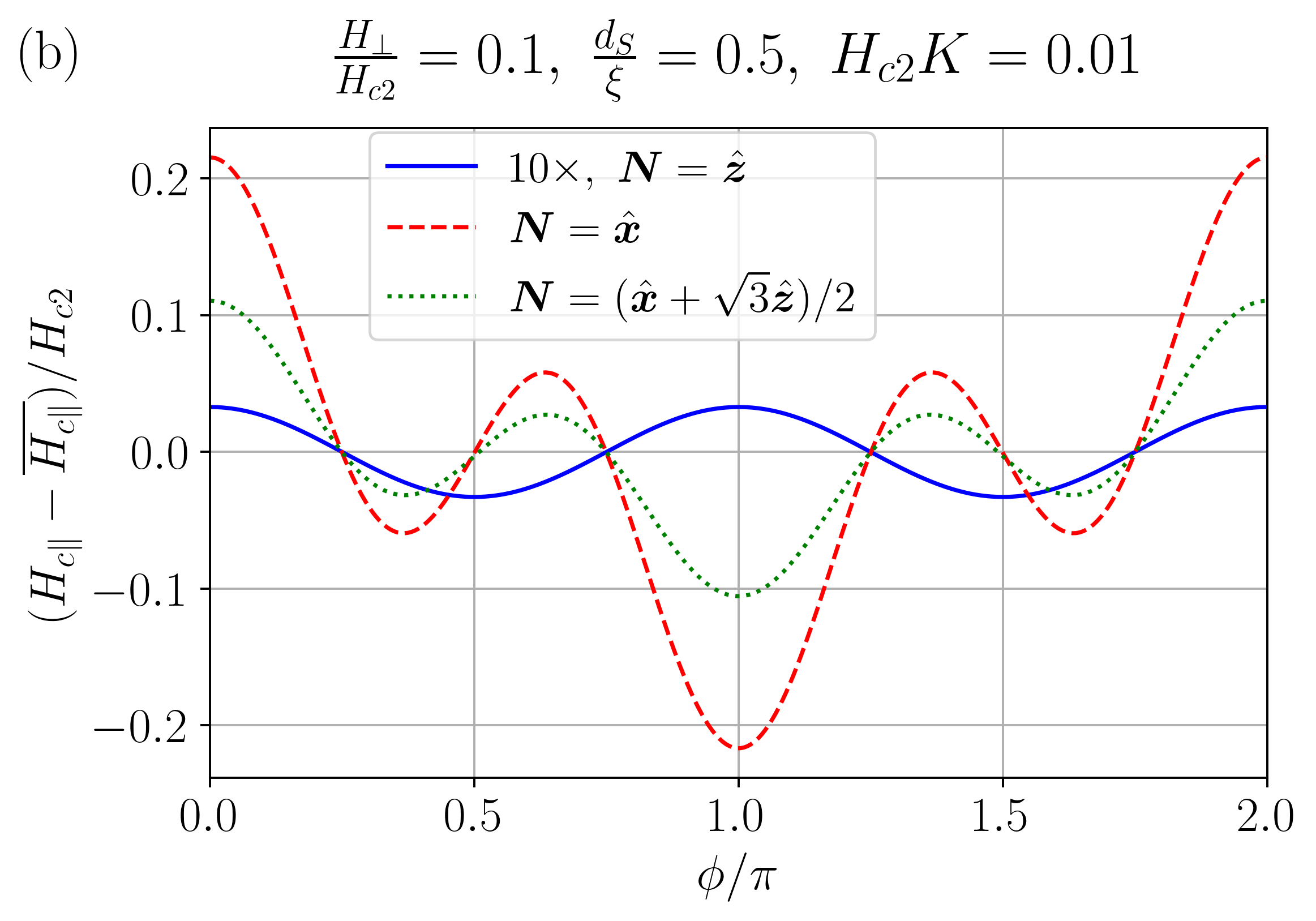}
    \includegraphics[width=0.92\linewidth]{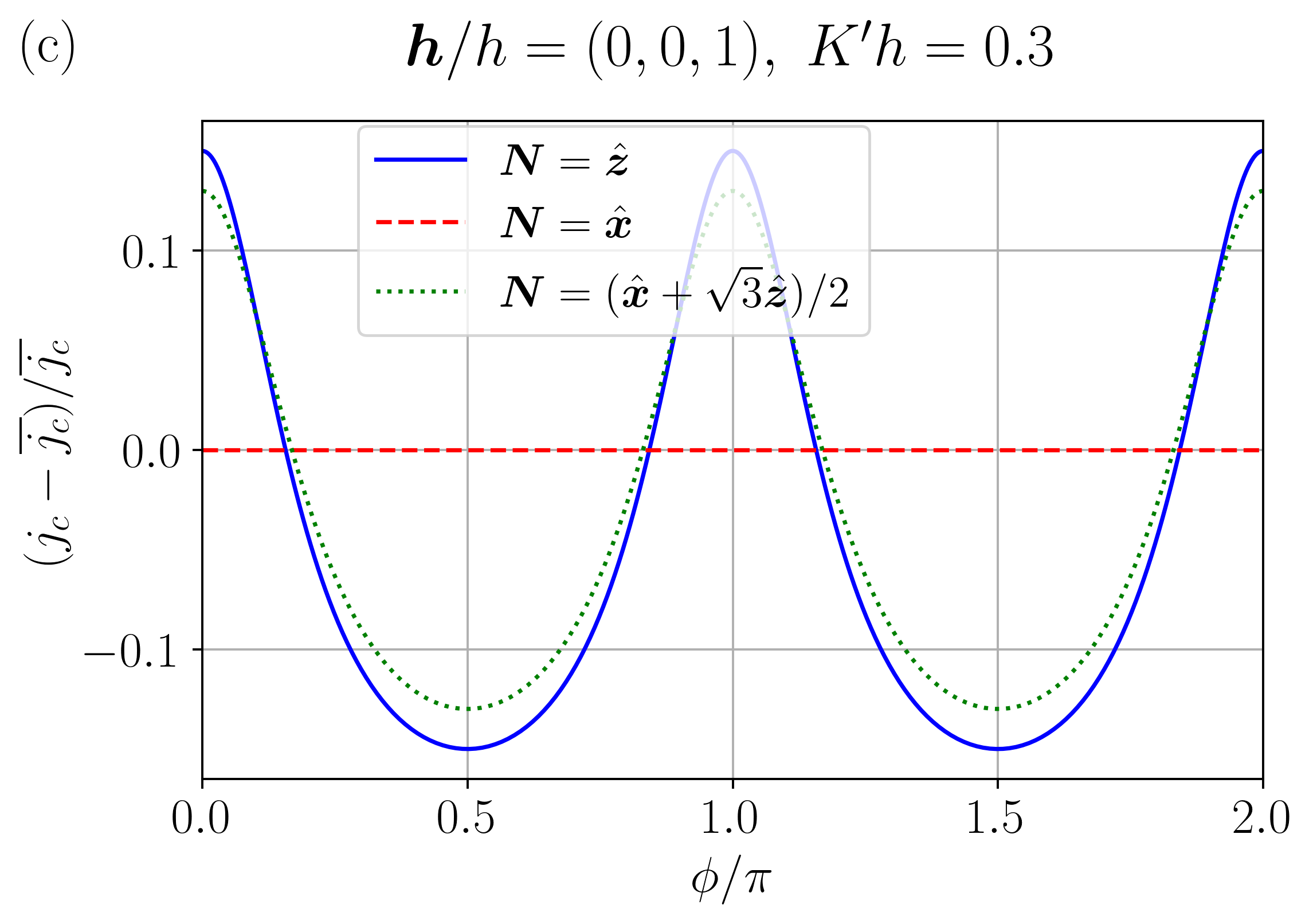}
    \caption{
    (a), (b): Critical temperature and parallel critical field for the setup shown in Fig.~\ref{fig:setup}(a), obtained from Eqs.~\eqref{eq:Tc} and \eqref{eq:Hc_simple}, respectively.  Here, $\xi^{(0)} = \xi(T=0)$ and $H_{c2}^{(0)} = H_{c2}(T=0)$, with
$\xi^2 = [2m^* \alpha |T_{c0}-T|]^{-1}$ and $H_{c2} = \Phi_0/(2\pi \xi^2)$. (c): Critical current density modulation for the setup shown in Fig.~\ref{fig:setup}(b), according to Eqs.~\eqref{eq:j_c_theta} and \eqref{eq:phi_theta}.  Here, the notation $\overline{q}$ is used to denote averaging of $q$ over the angle $\phi$ between  applied parallel magnetic field or supercurrent and one of the crystallographic axes with maximal spin-splitting.}
    \label{fig:observables}
\end{figure}

Motivated by these developments, in this work we address the question of whether superconductivity can provide complementary and experimentally accessible signatures of altermagnetic order, either in materials where the two orders coexist or in hybrid structures combining conventional superconductors with altermagnetic insulators. Focusing on collinear $d$-wave altermagnetic superconductors in an external magnetic field, we demonstrate that the three central superconducting observables -- the critical temperature, the parallel critical magnetic field, and the critical current density -- acquire anisotropies attributed to the interplay between superconductivity and altermagnetic ordering. Crucially, the magnitude and symmetry of these anisotropies are controlled by the relative angle between  the applied  magnetic field and the Néel vector, thereby providing a symmetry-resolved probe of the underlying altermagnetic order.  Our main results are summarized in Fig.~\ref{fig:observables}. 

We start from considering a thin film lying in the $\hat{\vec{x}}$--$\hat{\vec{y}}$ plane that hosts coexisting superconducting and collinear $d$-wave altermagnetic orders. These orders may coexist intrinsically within a single material, or the altermagnetic order may be induced in the superconducting film via the proximity effect from an altermagnetic insulator, as illustrated in Fig.~\ref{fig:setup}(a) \cite{vasiakin2025disorder,Belzig_2025_Thermodynamic,heras2025interplay,bobkov2026inverse}.  The film is subjected to an external magnetic field $\vec{H} = (H_\parallel \cos\phi, H_\parallel \sin\phi, H_\perp)$. Its dimensions are assumed to satisfy $\xi \ll d_x, d_y\ll \lambda_L$ and $d_S \ll \xi$, where  $d_x$ and $d_y$ are the film length and width along the $\hat{\vec{x}}$ and $\hat{\vec{y}}$ axes, respectively, $\xi$ is the coherence length and $\lambda_L$ is the London penetration depth, so the field penetrates fully into the sample and we may write $\vec{H} \approx \vec{B} = \operatorname{rot}\vec{A}$ where $\vec{A}$ is the vector potential.

To describe our superconducting films, we employ the Ginzburg–Landau (GL) approach. The corresponding  free energy functional, valid  up to the second order in the superconducting order parameter $\Psi$ has been derived in \cite{Zyuzin2024,heras2025interplay}. Since we are interested not only in the critical temperature and field, but also in the critical current of the film, we employ the GL functional up to the fourth order in $\Psi$ which reads as    
\begin{align}
    &F[\Psi] = \int dV\ \mathcal{F} = \int dV\,\left\{
 a \vert \Psi \vert^2
+ \frac{b}{2} \vert \Psi \vert^4 + \frac{H^2}{8\pi}    \right. \nonumber\\
 &\ \ \left.
 + \frac{1}{2m^*} \left( D_k \Psi\, D_k^\star \Psi^\star+H_a N_a  K_{jk}  D_j  \Psi\, D_k^\star \Psi^\star \right) \right\}, \label{eq:GL_Free_Energy}
\end{align}
where the integral extends over the entire sample.  Here, $a$, $b$, and $m^*$ represent the GL coefficients. The vector potential $\vec{A}$   defines the magnetic field via $\vec{H} = \nabla \times \vec{A}$, and enters the gauge-covariant derivatives as $D_k=-i\nabla_k-\frac{e^*}{c}A_k$, where $e^*=2e$ is the Cooper pair charge and $c$ is the speed of light. 
The term $H_a N_a K_{jk} D_j \Psi D_k^{\star} \Psi^{\star}$ in Eq.~\eqref{eq:GL_Free_Energy} describes the altermagnetism-induced correction to the conventional kinetic energy. Its the only scalar one can build from the time-reversal-odd third-rank tensor $N_a K_{jk}$. 

This tensor factorizes into the unit Néel vector $\vec N$ and a second-rank tensor $K_{jk}$ that encodes the symmetry of the altermagnetic spin splitting. Under time reversal, $\vec N$ is odd, while $K_{jk}$ is even and symmetric in its indices~\cite{Kokkeler_2025}.
Physically, this new term modifies the kinetic energy of the condensate by introducing an effective anisotropic electron mass. The degree of anisotropy is controlled by the projection of the external magnetic field onto the Néel vector $N_a H_a$. This behavior clearly distinguishes altermagnetic superconductors from conventional anisotropic superconductors~\cite{Larkin1992,Walker_1959_BCS_overlapping_bands,Bhandari_1972_SC_Magnetoresistance}, where the effective mass tensor does not depend on the external magnetic field. Without loss of generality, we choose the in-plane axes in Fig.~\ref{fig:setup}(a) such that the altermagnetic tensor has only two nonzero components, $K_{xx} = - K_{yy} = K$. This choice corresponds to the maximal spin splitting along the $\hat{\vec{x}}$ and $\hat{\vec{y}}$ directions in momentum space \cite{heras2025interplay}.
We notice that the form of the GL free energy,  Eq.~\eqref{eq:GL_Free_Energy}, is general and valid for an arbitrary degree of elastic disorder \cite{Zyuzin2024,heras2025interplay}. 

We begin exploring the functional Eq.~\eqref{eq:GL_Free_Energy} by calculating the critical temperature and the parallel critical  field of the film shown in Fig.~\ref{fig:setup}(a). 
Superconductivity in thin films 
is destroyed by increasing temperature or magnetic field through a second-order phase transition~\cite{tinkham2004introduction}. 
Close to the transition, $\vert \Psi \vert$ is small  and one can  obtain the linearized GL equation by taking the functional derivative of Eq.~\eqref{eq:GL_Free_Energy} with respect to $\Psi^\star$, retaining terms up to the first order in $\Psi$, and setting the result equal to zero. This yields:
\begin{align}
    \label{eq:linearizedGL}
    &\Bigg[ \frac{1+K N_a H_a}{2m^*}D_x^2+\frac{1-K N_a H_a}{2m^*} D_y^2 + \nonumber \\  
    &\qquad \qquad + \frac{1}{2m^*}D_z^2 \Bigg] \Psi= \vert a \vert \Psi.
\end{align}
The physical meaning of~\eqref{eq:linearizedGL} is straightforward: for $N_a H_a \neq  0$, the coefficients in front of $D_x^2$ and $D_y^2$ become unequal, generating an anisotropic field-dependent effective mass in the film plane.   Eq.~\eqref{eq:linearizedGL} is supplemented with boundary conditions suitable for the nucleation 
of superconducting phase problem, $\left. D_z \Psi \right\vert_{z = \pm d_S/2} = 0$ and $\Psi \to 0$ when $\sqrt{x^2 + y^2} \to \infty$, and with the vector potential $\vec{A} = (z H_{\parallel}\sin \phi, x H_\perp -z H_{\parallel}\cos \phi,0)$ leading to the magnetic field $\vec{H} = (H_\parallel \cos\phi, H_\parallel \sin\phi, H_\perp)$.

After averaging Eq.~\eqref{eq:linearizedGL} over the film width $d_S \ll \xi$, the problem of superconducting phase nucleation is analogous to the quantum mechanics of a harmonic oscillator \cite{tinkham2004introduction}. 
Using this analogy, we derive the critical temperature and the parallel critical field of the film up to the first order in $K N_a H_a$ (see the Appendix for details):
\begin{subequations}\label{eq:Tc_Hc}
\begin{align}
T_c - T_{c0}&= - \frac{e^* |H_\perp|}{2 m^* c \alpha} - \frac{e^{*2} d_S^2 H_{\parallel}^2}{24 m^* c^2 \alpha}
\left(1 - K N_a H_a \cos 2\phi \right),
\label{eq:Tc}
\\[4pt]
H_{c\parallel}/H_0 &=  1 + \frac{1}{2} K N_a \tilde{H}_a \cos 2\phi.
\label{eq:Hc_simple}
\end{align}
\end{subequations}
Here, $\tilde{\vec{H}} =  (H_0 \cos\phi, H_0\sin\phi, H_\perp)$, and $H_0 = 2\sqrt{3}  \frac{H_{c2} \xi}{d_S}\sqrt{1-\frac{\vert H_\perp \vert}{H_{c2}}}$ is the parallel critical field for a film without altermagnetic order, in which $H_{c2} = \Phi_0/(2\pi \xi^2)$ is the second critical field,  $\Phi_0 = \pi c/e$ is the flux quantum, and $\xi = 1/\sqrt{2m^* \vert T_{c0} - T\vert}$ is the GL coherence length.

The angular dependencies given by Eqs.~\eqref{eq:Tc} and \eqref{eq:Hc_simple} are shown in Figs.~\ref{fig:observables}(a) and (b). 
For $\vec{N}$ perpendicular to the film, both $T_c(\phi)$ and $H_{c\parallel}(\phi)$ are symmetric under $\phi \to \phi + \pi$ rotations. 
In contrast, when $\vec{N}$ lies in the $\hat{\vec{x}}$--$\hat{\vec{y}}$ plane, additional extrema appear, and the $\pi$-rotation symmetry is broken. 
If $\vec{N}$ has both in-plane and out-of-plane components, these additional extrema become less pronounced and gradually vanish as $\vec{N} \to \hat{\vec{z}}$.

From Eqs.~\eqref{eq:Tc} and \eqref{eq:Hc_simple}, we conclude that measurements of $T_c$ and $H_{c\parallel}$ as functions of the magnitude and direction of the external magnetic field $\vec{H}$ enable the extraction of the altermagnetic parameters $K$ and $N_a$. For instance, measuring $T_c(H_\perp,\vec{H}_\parallel)$ allows one to determine $K$ and $\vec{N}$ from the relations
\begin{subequations} \label{eq:Tc_mes}
\begin{align}
T_c(H_\perp,\vec{H}_\parallel) - T_c(H_\perp,-\vec{H}_\parallel)
&\propto H_\parallel^2 K\, \vec{N}_\parallel \cdot \vec{H}_\parallel,\label{eq:T_cpar}\\[4pt]
T_c(H_\perp,\vec{H}_\parallel) - T_c(-H_\perp,\vec{H}_\parallel)
&\propto H_\parallel^2 K\, N_\perp H_\perp. \label{eq:T_cperp}
\end{align}
\end{subequations}
Here, $N_\perp$ and $\vec{N}_\parallel$ are defined as  $N_\perp = \vec{N}\cdot \hat{\vec{z}} $ and $\vec{N}_\parallel = \vec{N} - \hat{\vec{z}} N_\perp$ Analogous relations apply to measurements of $H_{c\parallel}$.

Another experimental observable that may reflect the underlying altermagnetic $d$-wave symmetry is the dependence of the critical current on its direction and on the applied magnetic field $\vec{H}$, as we discuss next. In order to avoid dealing with a partial differential equation in two dimensions, we simplify the problem by assuming first that the magnetic field coupled to the Néel vector arises via exchange coupling from an adjacent layer of a ferromagnetic insulator (FI) \cite{Strambini_2017_MagneticProximity,Hijano_2021_CoexistenceBilayers,bobkov2026inverse}, see  Fig.~\ref{fig:setup}(b). In this situation, and because $d_S \ll \xi$, both the exchange field $h$ induced by the FI layer and the altermagnetic order owing to the AMI layer, characterized by the Néel vector $\vec{N}$, coexist within the superconductor, leading to an effective coupling between them. In other words, the functional in Eq.~\eqref{eq:GL_Free_Energy} preserves its form upon making the substitution $N_a H_a K_{ij} D_i \Psi D^\star_j \Psi^\star \to N_a h_a K'_{ij} D_i \Psi D^\star_j \Psi^\star$. The constants $K$ and $K'$ are related by $K' = K/(g \mu_B)$, where $g$ is the electron g-factor and $\mu_B$ is the Bohr magneton. 

To  compute $j_c$ in the thin film, Fig.~\ref{fig:setup}(b), we use Bardeen's approach~\cite{Bardeen1962,schmidt2013physics}. The magnitude of the order parameter, $\vert \Psi \vert^2$, can be assumed to be constant along the width of the film due to the condition $d_S \ll \xi$, and a uniform phase gradient of the order parameter is imposed across the film:
\begin{equation} \label{eq:p_S_given}
    \vec{q}_S = \nabla \varphi 
    = q_S(\cos\theta, \sin\theta, 0),
\end{equation}
where $\varphi$ is the phase of the condensate wavefunction, $\Psi = \vert \Psi \vert e^{i\varphi}$. The current density is then obtained as:
\begin{equation}\label{eq:current}
    j_{Sk}=-\frac{\partial \mathcal{F}}{\partial A_k}=\frac{e^*}{m^*}\vert \Psi \vert^2(\delta_{jk}+ N_a h_a K'_{jk})q_{Sj}.
\end{equation}
The presence of the altermagnetic tensor $K'_{jk}$ modifies the current–momentum relation, so that the supercurrent is generally not collinear with the phase gradient, $\vec{j} \nparallel \vec{q}_S$ \cite{Zyuzin2024}. This non-collinearity is analogous to the one reported in  anisotropic superconductors, where an anisotropic effective mass tensor similarly leads to a misalignment between supercurrent and phase gradient \cite{Bhandari_1972_SC_Magnetoresistance,abrikosov2017fundamentals}. The main difference here is that in our case the renormalized mass depends on the applied field.

We minimize the free energy of the film given by Eq.~\eqref{eq:GL_Free_Energy} with respect to $\vert \Psi \vert^2$ considering that $\vert \Psi \vert^2$ and $\vec{q}_S$ are constant along the film width $d_S \ll \xi$ and find  the $\vert \Psi \vert^2(\vec{q}_S)$ dependence. Substituting this result into Eq.~\eqref{eq:current}, we determine the critical current as $j_c = \max_{q_S} \sqrt{ j_{Sk}(q_S)\, j_{Sk}(q_S) }$. This procedure gives the following result valid up to the first order in $K' h$:
\begin{equation}\label{eq:j_c_theta}
    j_c = \frac{2^{3/2}e^*|a|^{3/2}}{3^{3/2}m^{*1/2}b}\left(1+ \frac{1}{2}K' N_a h_a \cos2\theta \right).
\end{equation}
In this equation, the angle $\theta$ defines the direction of the imposed phase gradient $\vec q_S$  with respect to one of the crystallographic axes of the film corresponding to the maximum spin-splitting. In order to determine the direction of the supercurrent $\vec{j}_S = j_S(\cos\phi, \sin\phi, 0)$ created by the phase gradient $\vec{q}_S = q_S(\cos\theta,\sin\theta,0)$, we employ the supercurrent definition Eq.~\eqref{eq:current} and obtain the following connection between the angles $\phi$ and $\theta$
\begin{equation} \label{eq:phi_theta}
    \tan \phi = \frac{1 - K' N_a h_a}{1 + K' N_a h_a} \tan \theta.
\end{equation}
We solve Eq.~\eqref{eq:phi_theta} for the dependence $\theta(\phi)$, substitute it into Eq.~\eqref{eq:j_c_theta} and obtain the critical current density $j_c$ as a function of the exchange field $\vec{h}$ and the direction of the applied supercurrent given by $\phi$. In the limit $K' N_a h_a \ll 1$,  $\phi \approx \theta$, and  the critical current $j_c(\vec{h},\phi)$ satisfies relations analogous to Eqs.~(\ref{eq:T_cpar}-\ref{eq:T_cperp}), namely $j_c(h_\perp,\vec{h}_\parallel,\phi) - j_c(h_\perp,-\vec{h}_\parallel,\phi)
\propto K' \,\vec{N}_\parallel \cdot \vec{h}_\parallel$, $j_c(h_\perp,\vec{h}_\parallel,\phi) 
- j_c(-h_\perp,\vec{h}_\parallel,\phi)
\propto K' \, N_\perp h_\perp$.

The dependence $j_c(\phi)$, shown in Fig.~\ref{fig:observables}(c), clearly shows the emergence of $d$-wave modulation of the critical current density  of the superconducting film. At first glance, this dependence may suggest that the application of the exchange field $\vec h$ enhances the critical current along certain directions. However, the usual GL coefficients are renormalized by the altermagnetic order, as demonstrated in \cite{heras2025interplay}. As a result, the zero-field critical current of an altermagnetic superconductor is intrinsically reduced compared to that of a conventional BCS superconductor. The apparent field-induced enhancement in Eq.~\eqref{eq:j_c_theta} should be understood relative to this already suppressed altermagnetic baseline, rather than as an absolute increase beyond the conventional superconducting limit. The same argument applies to the other observables analyzed in this work.

The above results are obtained by assuming a pure exchange coupling stemming from the FI film, in Fig.~\ref{fig:setup}(b). This setup may be experimentally challenging, and therefore we propose an alternative setup, shown in Fig.~\ref{fig:setup}(c). It consists of a  cross-shaped superconducting layer on top of the altermagnetic insulator. 

The two  independent transport channels that support supercurrent flow are oriented along the $\hat{\vec{x}}$ and $\hat{\vec{y}}$ axes.
Each channel is assumed to be narrow $d_y, d_x \ll \xi$, in addition to the thin-film condition $d_S \ll \xi$. Thus, we neglect the depairing effect of $H_\parallel$. In this case, the problem can be treated separately depending on the current direction, so the vector potential can be chosen as $\vec{A} = (-y H_\perp, 0, 0)$ for a channel along $\hat{\vec{x}}$ and as $\vec{A} = (0, x H_\perp, 0)$ for a channel along $\hat{\vec{y}}$.

We again employ the Bardeen approach, which involves  the gradient of the order parameter phase $\vec{q}_S = \nabla \varphi$ along the channels,  two successive spatial averagings of the GL equation obtained from the functional derivative of Eq.~\eqref{eq:GL_Free_Energy}:  over the film thickness $d_S$ (taking $\vec{q}_S$ being constant along the width), and then over the coordinate transverse to the supercurrent flow in each channel. 
We thereby determine the Cooper-pair density $\vert \Psi \vert^2$ as a function of the applied phase gradient, substitute it into Eq.~\eqref{eq:current}, perform the integrations along the current transverse coordinates, and search for the maxima of $I^2_{x}$ and $I^2_{y}$. After this procedure, we obtain the following critical currents for the two transport channels:
\begin{subequations}\label{eq:Icx_Icy}
\begin{align}
    &I_{cx} = \frac{2 e^* \vert a \vert d_y d_S }{3^{3/2} m^* b \xi} \left(1 - \frac{1}{8}\left(\frac{ H_\perp d_y}{H_{c2} 
    \xi}\right)^2 + m^* K H_a N_a\right),\label{eq:Icx}\\[4pt]
    &I_{cy} = \frac{2  e^* \vert a \vert  d_x d_S }{3^{3/2} m^* b \xi} \left(1 - \frac{1}{8}\left(\frac{H_\perp d_x }{H_{c2} \xi}\right)^2 - m^* K H_a N_a\right). \label{eq:Icy}
\end{align}
\end{subequations}
From Eqs.~\eqref{eq:Icx} and \eqref{eq:Icy}, we find that in the cross-shaped geometry
$I_{cx}(\vec{H}) - I_{cy}(\vec{H}) \propto  K N_a H_a$.  Thus, measurement of the difference between the critical currents in the two directions provides another direct probe of the altermagnetic parameters $K$ and $\vec{N}$.

In conclusion, we propose a straightforward and experimentally simple method to identify collinear $d$-wave altermagnetic ordering through measurements of the critical properties of an adjacent superconductor. Our predictions are particularly well suited for altermagnetic insulators (AMI) with a thin superconducting (SC) layer grown on top. Specifically, we derive simple analytic expressions that can be readily used by experimentalists for the critical temperature, in-plane critical field, and critical current as functions of externally controlled parameters, such as the direction of the applied magnetic field and the current direction. From these dependencies, one can extract information about the altermagnetic coupling $K$ and the Néel vector $\vec{N}$. From the materials perspective, the AMI can be chosen from the proposed $d$-wave altermagnets listed, for example, in Ref.~\cite{Smejkal2022Emerging}, or from antiferromagnets with surface-emergent altermagnetic order recently predicted in Ref.~\cite{lange2026emergent}. For the superconducting layer, any conventional superconductor may be used, such as Al or Nb.

{\it Acknowledgments.}
We thank Tim Kokkeler and Ilya Tokatly for useful discussions. R.~H. and F.~S.~B. thank financial support from the the Spanish MCIN/AEI/10.13039/501100011033 
through the grants PID2023-148225NB-C31, and TED2021-130292B-C41. A. A. M. and F. S. B. also thank financial support from the
European Union’s Horizon Europe research and innovation program under grant agreement No. 101130224 (JOSEPHINE). 


\bibliographystyle{apsrev4-1}
\bibliography{refs}
\clearpage
\onecolumngrid
\appendix

\section*{Derivation of \texorpdfstring{$T_c$}{Tc} and \texorpdfstring{$H_{c\parallel}$}{Hc parallel}}
\label{sec:Tc_Hc_derivation}

Starting from Eq.~\eqref{eq:linearizedGL} of the main text, we derive the critical temperature $T_c(\vec H)$ and the parallel critical field $H_{c\parallel}(H_\perp, \phi)$ following the standard procedure described, for example, in Ref.~\cite{tinkham2004introduction}. The boundary conditions complementing Eq.~\eqref{eq:linearizedGL} are:
\begin{equation}\label{eq:bcond}
    \left.D_z\Psi\right|_{z=\pm d_S/2}=0,\qquad \Psi \to 0\text{, at } \sqrt{x^2 + y^2} \to \infty,
\end{equation} 
which correspond to the nucleation of a superconducting phase in the film. We choose the gauge:
\begin{equation}
    \vec A=(z H_{\parallel}\sin \phi,xH_\perp - zH_{\parallel}\cos\phi,0).
\end{equation}
With this choice, the system is translationally invariant along the $y$ direction, and the condensate wavefunction can be written as $\Psi(x,y,z)=e^{iq_{Sy}y}f(x,z)$. Since the critical field (or critical temperature) is determined by the lowest eigenvalue of Eq.~\eqref{eq:linearizedGL}, we set $q_{Sy} = 0$. Then Eq.~\eqref{eq:linearizedGL} becomes:
\begin{equation} \label{eq:A3}
    \left[-\frac{1+K N_a H_a}{2m^*}\left(\nabla_x-\frac{i e^*}{c}z H_{\parallel}\sin \phi \right)^2+\frac{1-KN_a H_a}{2m^*}\frac{e^{*2}}{c^2}\left(x H_\perp -z H_{\parallel}\cos\phi \right)^2-\frac{1}{2m^*}\nabla_z^2\right]f=|a|f.
\end{equation}
In the thin film limit $d_S\ll\xi$, the order parameter varies weakly along $z$, and we may approximate $f(x,z) \approx f(x)$, which automatically satisfies the first boundary condition in Eq.~\eqref{eq:bcond}. We  replace $z$ and $z^2$ by their averages across the film thickness $\langle z\rangle=0$ and $\langle z^2 \rangle=\frac{d^2_S}{12}$. Eq.~\eqref{eq:A3} is then reduced to
\begin{equation}
    \Bigg[-\frac{1}{2m^*}\nabla_x^2+\frac{1}{2}m^*\frac{1-K N_a H_a }{1+K N_a H_a}\frac{e^{*2}H_\perp^2}{m^{*2}c^2}x^2\Bigg] f=\frac{1}{1+K N_a H_a }\Bigg[ \vert a \vert-\frac{e^{*2}d_S^2H_{\parallel}^2}{24m^*c^2}(1-KN_a H_a \cos 2\phi)\Bigg]f.
\end{equation}
This equation, together with the boundary conditions on $f$ following from Eq.~\eqref{eq:bcond}, are analogous to the problem of a quantum harmonic oscillator with frequency $\omega_c=\sqrt{\frac{1-K N_a H_a}{1+K N_a H_a}}\frac{e^{*} \vert H_\perp \vert }{m^{*}c}$, 
whose lowest eigenvalue is $E_0=\omega_c/2$. 
This correspondence yields the following:
\begin{align}\label{eq:lowesteigenvalue}
    E_0 = \frac{\omega_c}{2} &=  \frac{1}{1+K N_a H_a }\Bigg[ \vert a \vert-\frac{e^{*2}d_S^2H_{\parallel}^2}{24m^*c^2}(1-KN_a H_a \cos 2\phi)\Bigg],\\  
    \vert a \vert &=\frac{1}{2}(1 + KN_a H_a)\omega_c +\frac{e^{*2}d_S^2H_{\parallel}^2}{24m^*c^2}(1-K N_a H_a \cos 2\phi) = \nonumber\\
    &= \frac{e^{*2}d_S^2H_{\parallel}^2}{24m^*c^2}(1-K N_a H_a \cos 2\phi)+\frac{e^*|H_\perp|}{2m^*c}+O\left( (K N_a H_a)^2 \right),
\end{align}
where we only keep terms of linear order in $KN_a H_a$ in the last step. Since we are interested in the critical temperature $T_c$ for a given magnetic field $\vec{H}$, we substitute the definition of $\vert a\vert$ in the form  $ \vert a \vert =\alpha(T_{c0}-T_c)$ in which $T_{c0}$ is the zero-field critical temperature and $\alpha>0$, in Eq.~\eqref{eq:lowesteigenvalue} and obtain:
\begin{equation}
    T_c =  T_{c0}-\frac{e^*|H_\perp|}{2m^*c\alpha}-\frac{e^{*2}d_S^2H_{\parallel}^2}{24m^*c^2\alpha}\left(1-K N_a H_a \cos 2\phi \right) +O\left( (K N_a H_a)^2 \right),
\end{equation}
which corresponds to Eq.~\eqref{eq:Tc} of the main text.

In order to compute the parallel critical field $H_{c\parallel}$ at a given temperature $T$, we define the coherence length as $\xi^2=\frac{1}{2m^* \vert a \vert}$ where $a = \alpha (T_{c0} - T)$, the flux quantum $\Phi_0=\frac{\pi c}{e}$, and choose the following representation of the Néel vector, $\vec N=(N_{\parallel}\cos \alpha ,N_{\parallel}\sin \alpha,N_\perp)$. With the help of these definitions, we rewrite Eq.~\eqref{eq:lowesteigenvalue} retaining only linear terms with respect to $K N_a H_a$ and find the following equation on $H_{c\parallel}$:
\begin{equation}\label{eq:auxHcparallel}
    H_{c\parallel}^2 \left(1-K N_a H_a \cos 2\phi \right)=\frac{3\Phi_0^2}{\pi^2\xi^2d_S^2}\left(1-\frac{2\pi\xi^2|H_\perp|}{\Phi_0}\right).
\end{equation}
We define $H_0=\frac{\sqrt{3}\Phi_0}{\pi\xi d_S}\sqrt{1-\frac{2\pi \xi^2|H_\perp|}{\Phi_0}}$ and rewrite Eq.~\eqref{eq:auxHcparallel} which now reads:
\begin{equation} \label{eq:Hc_eq_appendix}
    -K\cos(\alpha-\phi)N_{\parallel}H_{c\parallel}^3 \cos 2\phi +\left(1-KN_\perp H_\perp\cos 2\phi\right)H_{c\parallel}^2=H_0^2.
\end{equation}
Solving this equation perturbatively up to the linear order in $K N_a H_a$ yields two solutions:
\begin{equation}
    H_{c\parallel}=\pm H_0\left[1+\frac{K}{2}(N_\perp H_\perp\pm N_{\parallel}H_0\cos(\alpha-\phi)) \cos 2\phi \right].
\end{equation}
Since $H_{c\parallel}(K=0) > 0$ \cite{tinkham2004introduction}, we retain only the positive branch:
\begin{equation} 
    H_{c\parallel}= H_0\left[1+\frac{K}{2}(N_\perp H_\perp+ N_{\parallel}H_0\cos(\alpha-\phi)) \cos 2\phi\right],
\end{equation}
which corresponds to Eq.~\eqref{eq:Hc_simple} of the main text.

\end{document}